\documentclass[11pt]{article}

\usepackage[utf8]{inputenc}
\usepackage{amsmath}
\usepackage{amssymb}
\usepackage{amsthm}
\usepackage{color,soul}
\usepackage{tikz}
\usetikzlibrary{topaths,calc}
\usepackage{subfigure}
\usepackage{cite}

\def\QED{\mbox{\rule[0pt]{1.5ex}{1.5ex}}}
\def\proof{\noindent\hspace{2em}{\it Proof: }}

\def\QED{\mbox{\rule[0pt]{1.5ex}{1.5ex}}}

\newtheorem{theorem}{Theorem}

\newtheorem{lemma}{Lemma}

\newcommand\blfootnote[1]{%
  \begingroup
  \renewcommand\thefootnote{}\footnote{#1}%
  \addtocounter{footnote}{-1}%
  \endgroup
}

\title{Secure Aggregation with an Oblivious Server}
\author{Hua Sun}
\date{}

\begin{document}
\maketitle

\blfootnote{
Hua Sun (email: hua.sun@unt.edu) is with the Department of Electrical Engineering at the University of North Texas.}

\begin{abstract}
Secure aggregation usually aims at securely computing the sum of the inputs from $K$ users at a server. Noticing that the sum might inevitably reveal information about the inputs (when the inputs are non-uniform) and typically the users (not the server) desire the sum (in applications such as federated learning), we consider a variant of secure aggregation where the server is oblivious, i.e., the server only serves as a communication facilitator/helper to enable the users to securely compute the sum and learns nothing in the process. Our communication protocol involves one round of messages from the users to the server and one round of messages from the server to each user such that in the end each user only learns the sum of all $K$ inputs and the server learns no information about the inputs. For this secure aggregation with an oblivious server problem, we show that to compute $1$ bit of  the sum securely, each user needs to send at least $1$ bit to the server, the server needs to send at least $1$ bit to each user, each user needs to hold a key of at least $2$ bits, and all users need to collectively hold at least $K$ key bits. In addition, when user dropouts are allowed, the optimal performance remains the same, except that the minimum size of the key held by each user increases to $K$ bits, per sum bit.
\end{abstract}

\newpage
\allowdisplaybreaks
\section{Introduction}
\begin{figure}[h]
\centering
\begin{tikzpicture}
    \node (u1) at (0,4.1) {};
    \node (u2) at (0,2.5) {};
    \node at (0.5,1.7) {$\vdots$};
    \node (uK) at (0,0) {};
    \node (server) at (8,2.25) {};
    \node (server1) at ($(server)+(0.1,-0.5)$) {};
    \node (server2) at ($(server)+(0.1,-0.2)$) {};
    \node (server3) at ($(server)+(0.1,0.1)$) {};
    \node (server4) at ($(server)+(0.1,0.4)$) {};
    \filldraw ($(u1)$)
    to[out=90,in=180] ($(u1) + (0.5,0.5)$)
    to[out=0,in=90] ($(u1) + (1,0)$);
    \fill ($(u1) + (0.5,0.8)$) circle(0.3);
    \filldraw ($(u2)$)
    to[out=90,in=180] ($(u2) + (0.5,0.5)$)
    to[out=0,in=90] ($(u2) + (1,0)$);
    \fill ($(u2) + (0.5,0.8)$) circle(0.3);
    \filldraw ($(uK)$)
    to[out=90,in=180] ($(uK) + (0.5,0.5)$)
    to[out=0,in=90] ($(uK) + (1,0)$);
    \fill ($(uK) + (0.5,0.8)$) circle(0.3);
    \filldraw ($(server)+(0,-0.6)$) rectangle ($(server)+(1,0.7)$);
    \foreach \v in {1,2,...,4} {
        \filldraw [white] (server\v) rectangle ($(server\v)+(0.8,0.2)$);
        \filldraw ($(server\v)+(0.3,0.08)$) rectangle ($(server\v)+(0.75,0.12)$);
        \fill ($(server\v)+(0.15,0.1)$) circle (0.05);
    }
    \draw [rounded corners,->, line width=1pt,shorten >=10pt]($(u1) + (1.15,0.2)$) -- ($(u1) + (5.5,0.2)$)
    -- (server);
    \draw [rounded corners,->, line width=1pt,shorten >=10pt]($(u2) + (1.15,0.2)$) -- ($(u2) + (5.5,0.2)$)
    -- (server);
    \draw [rounded corners,->, line width=1pt,shorten >=10pt]($(uK) + (1.15,0.2)$) -- ($(uK) + (5.5,0.2)$)
    -- (server);
    \node at ($(u1) + (1,0.5)$) [right]{$X_1=W_1+N_1$};
    \node at ($(u2) + (1,0.5)$) [right]{$X_2=W_2+N_2$};
    \node at ($(uK) + (1,0.5)$) [right]{$X_K=W_K-\sum_{k=1}^{K-1} N_k$};
    \node at ($(u1) + (0.5,-0.2)$) {User $1$};
    \node at ($(u2) + (0.5,-0.2)$) {User $2$};
    \node at ($(uK) + (0.5,-0.2)$) {User $K$};
    \node at ($(u1) + (0,0.5)$) [left]{$W_1,Z_1$};
    \node at ($(u2) + (0,0.5)$) [left]{$W_2,Z_2$};
    \node at ($(uK) + (0,0.5)$) [left]{$W_K,Z_K$};
    \node at ($(server)+(0.5,-0.9)$) []{Server};
    \node at ($(server)+(2.5,0.4)$) []{only learn};
    \node at ($(server)+(1,-0.05)$) [right]{$W_1+W_2+\cdots+W_K$};
    \node at ($(server)+(1.2,-0.5)$) [right]{$= X_1+\cdots+X_K$};
\end{tikzpicture}
\vspace{-0.1in}
\caption{Secure summation and an optimal protocol, where $N_1, \cdots, N_{K-1}$ are i.i.d. uniform over the same field as $W_k$ and $Z_k = N_k, k \in \{1,\cdots,K-1\}, Z_K = -\sum_{k=1}^{K-1} N_k$.}
\label{fig:sum}
\end{figure}
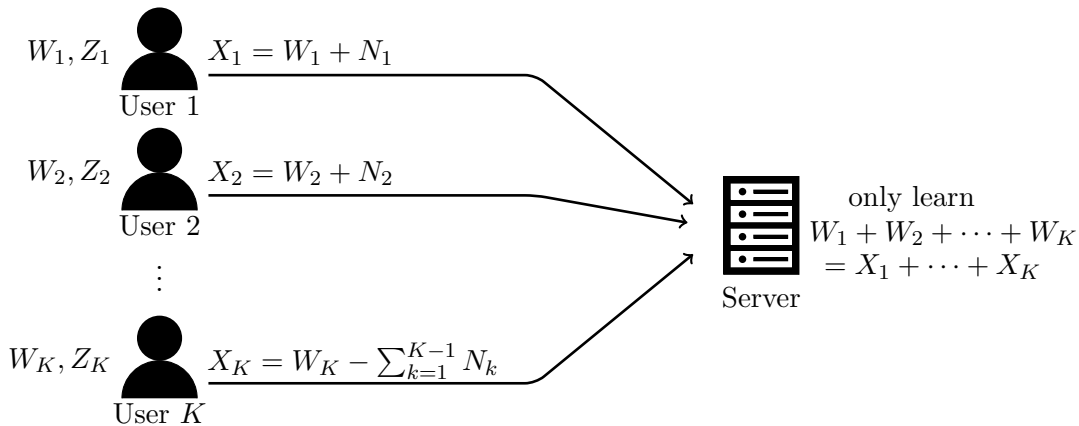

Secure aggregation \cite{aggregation, Zhao_Sun_Aggregate, aggregation_light, aggregation_swift, WSJC_Groupwise} arises in studying federated learning and is motivated by the need to securely compute the sum of gradients of distributed users without leaking information about individual gradients. Lying at the core is the secure summation problem \cite{Zhao_Sun_Sum} (see Fig.~\ref{fig:sum}), where $K$ users, with inputs (gradients) $W_k, k \in \{1,2,\cdots,K\}$ and keys $Z_k$, wish to compute and only compute the sum of $W_k$ at a server through orthogonal messages $X_k$. The new terminology, secure aggregation, is used to highlight certain new features brought by federated learning, e.g., user dropouts, that go beyond the basic secure summation problem. In this work, as user dropouts will be considered, for simplicity and consistency we will use the term secure aggregation (instead of both secure aggregation and secure summation) henceforth.

While individual inputs $W_k$ are `hidden' in the aggregated sum $\sum_{k=1}^K W_k$, the server might still obtain information about $W_k$ when the inputs are far from uniform. For example, suppose $K=3$, $W_1 \in \{0,1,2\}$, $W_2 \in \{0, 10, 20\}$, $W_3 \in \{0, 100, 200\}$ and the summation is over a large prime field. In this case, $\sum_{k=1}^3 W_k$ is invertible to $W_1, W_2, W_3$, i.e., all inputs are fully recovered from the sum and aggregation is not hiding anything. As a less extreme example, suppose $K=2$, $W_1 \in \{0,1\}$, $W_2 \in \{0,1\}$ and the summation is over $\mathbb{F}_3$. In this case, if $W_1 + W_2 = 2$, then we know for sure that $W_1 = W_2 = 1$ (the case where $W_1 + W_2 = 0$ is similar and when $W_1 + W_2 = 1$, we also know that $(W_1, W_2) = (0,1)$ or $(1,0)$). Therefore secure aggregation may not be as secure as we might grant even though information theoretic security is guaranteed. To make matters worse, if we may adversarially choose some inputs (e.g., through collusion), such leakage will be larger and ideas along this line have been applied in federated learning context to perform attacks to reveal information about the gradients and then about users' sensitive data \cite{zhao2023secure}. In this work, to make secure aggregation secure `again' (to regain our intuitive demand of security), we propose to change the role of the server from the entity that does sum computing to a pure helper, i.e., the server is not allowed to learn anything throughout and behaves only as an oblivious communication facilitator 
so that each user is able to compute the sum securely. Note that moving sum computation from the server to the users functions well in federated learning as the sum of gradients is required only at the user side to iteratively refine the model trained and the server does not necessarily need the gradient sum in the learning process. 

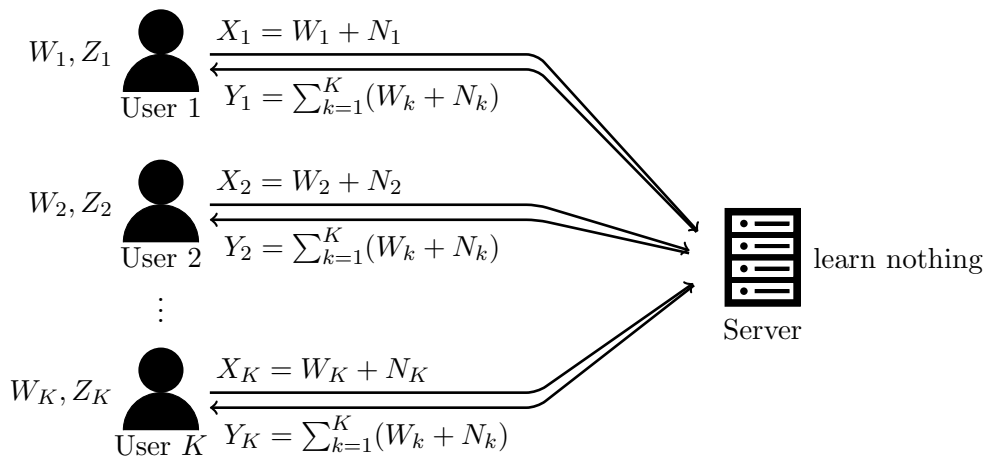
\begin{figure}[h]
\centering
\begin{tikzpicture}
    \node (u1) at (0,4.5) {};
    \node (u2) at (0,2.5) {};
    \node at (0.5,1.7) {$\vdots$};
    \node (uK) at (0,0) {};
    \node (server) at (8,2.25) {};
    \node (server1) at ($(server)+(0.1,-0.5)$) {};
    \node (server2) at ($(server)+(0.1,-0.2)$) {};
    \node (server3) at ($(server)+(0.1,0.1)$) {};
    \node (server4) at ($(server)+(0.1,0.4)$) {};
    \filldraw ($(u1)$)
    to[out=90,in=180] ($(u1) + (0.5,0.5)$)
    to[out=0,in=90] ($(u1) + (1,0)$);
    \fill ($(u1) + (0.5,0.8)$) circle(0.3);
    \filldraw ($(u2)$)
    to[out=90,in=180] ($(u2) + (0.5,0.5)$)
    to[out=0,in=90] ($(u2) + (1,0)$);
    \fill ($(u2) + (0.5,0.8)$) circle(0.3);
    \filldraw ($(uK)$)
    to[out=90,in=180] ($(uK) + (0.5,0.5)$)
    to[out=0,in=90] ($(uK) + (1,0)$);
    \fill ($(uK) + (0.5,0.8)$) circle(0.3);
    \filldraw ($(server)+(0,-0.6)$) rectangle ($(server)+(1,0.7)$);
    \foreach \v in {1,2,...,4} {
        \filldraw [white] (server\v) rectangle ($(server\v)+(0.8,0.2)$);
        \filldraw ($(server\v)+(0.3,0.08)$) rectangle ($(server\v)+(0.75,0.12)$);
        \fill ($(server\v)+(0.15,0.1)$) circle (0.05);
    }
    \draw [rounded corners,->, line width=1pt,shorten >=10pt]($(u1) + (1.15,0.5)$) -- ($(u1) + (5.5,0.5)$)
    -- (server);
    \draw [rounded corners,<-, line width=1pt,shorten >=10pt]($(u1) + (1.15,0.3)$) -- ($(u1) + (5.5,0.3)$)
    -- (server);
    \draw [rounded corners,->, line width=1pt,shorten >=10pt]($(u2) + (1.15,0.5)$) -- ($(u2) + (5.5,0.5)$)
    -- (server);
    \draw [rounded corners,<-, line width=1pt,shorten >=10pt]($(u2) + (1.15,0.3)$) -- ($(u2) + (5.5,0.3)$)
    -- (server);
    \draw [rounded corners,->, line width=1pt,shorten >=10pt]($(uK) + (1.15,0.5)$) -- ($(uK) + (5.5,0.5)$)
    -- (server);
    \draw [rounded corners,<-, line width=1pt,shorten >=10pt]($(uK) + (1.15,0.3)$) -- ($(uK) + (5.5,0.3)$)
    -- (server);
    \node at ($(u1) + (1.1,0.8)$) [right]{$X_1=W_1+N_1$};
    \node at ($(u1) + (1.2,-0.05)$) [right]{$Y_1=\sum_{k=1}^K (W_k+N_k)$};
    \node at ($(u2) + (1.1,0.8)$) [right]{$X_2=W_2+N_2$};
    \node at ($(u2) + (1.2,-0.05)$) [right]{$Y_2=\sum_{k=1}^K (W_k+N_k)$};
    \node at ($(uK) + (1.1,0.8)$) [right]{$X_K=W_K+N_K$};
    \node at ($(uK) + (1.2,-0.05)$) [right]{$Y_K=\sum_{k=1}^K (W_k+N_k)$};
    \node at ($(u1) + (0.5,-0.2)$) {User $1$};
    \node at ($(u2) + (0.5,-0.2)$) {User $2$};
    \node at ($(uK) + (0.5,-0.2)$) {User $K$};
    \node at ($(u1) + (0,0.5)$) [left]{$W_1,Z_1$};
    \node at ($(u2) + (0,0.5)$) [left]{$W_2,Z_2$};
    \node at ($(uK) + (0,0.5)$) [left]{$W_K,Z_K$};
    \node at ($(server)+(0.5,-0.9)$) []{Server};
    \node at ($(server)+(2.3,0)$) []{learn nothing};
\end{tikzpicture}
\vspace{-0.1in}
\caption{Secure aggregation with an oblivious server where the sum is securely computed at the users and an optimal protocol.}
\label{fig:oblivious}
\end{figure}

The specific communication model we consider is as follows and is probably the simplest one. Referring to Fig.~\ref{fig:oblivious}, the users first send a round of messages $X_k$ to the server and then the server replies back to the users with another round of messages $Y_k$. It turns out that the optimal protocol is straightforward, where the keys are set as $Z_k = (N_k, \sum_{k=1}^K N_k)$ and $N_1, \cdots, N_K$ are i.i.d. uniform over the same field as $W_k$. The messages $X_k, Y_k$ are set as $X_k = W_k + N_k$ and $Y_k = \sum_{k=1}^K X_k$, i.e., the noise variables in $X_k$ are i.i.d. and the server simply sends to each user the sum of $X_k$. Equipped with $\sum_{k=1}^K N_k$ as part of the key held by each user, everyone may recover and only recover $\sum_{k=1}^K W_k$ from $Y_k$. The main technical contribution of this work is to prove that the above natural protocol is information theoretically optimal in terms of both communication efficiency and key consumption. That is, to securely compute $1$ bit of the sum of the inputs, each message $X_k, Y_k$ needs to contain at least $1$ bit, each key $Z_k$ needs to contain at least $2$ bits, and all keys must have joint entropy of at least $K$ bits.

Introducing oblivious helpers in secure computation is by no means (in fact, far from) new. First, the generic model of secure multi-party computation is very general \cite{BGW, CCD, cramer_damgard_nielsen_2015}, where all users are allowed to have inputs and possibly distinct outputs (desired functions to compute). Hence an oblivious server is nothing but a party with no input and also no output requirements such that the resulting model is covered as a special case by general secure multi-party computation. Second, secure multi-party computation with oblivious helpers has been extensively studied as an explicit separate class of model in computer science and cryptography \cite{hirt2000player, cramer2005share, garay2017price, goyal2022tight, schneider2016lean} 
although the motivation and performance metric are quite different from ours. In light of the above discussion, our main intention is to introduce the oblivious helper model to 
secure aggregation (which is indeed natural and practical from the previous motivating paragraph) and to characterize the exact information theoretic limits on the communication and key rates.

Finally, returning to our model (see Fig.~\ref{fig:oblivious}), we include user dropouts (which is perhaps the most prominent new feature rooted in federated learning), where after sending $X_k$ to the servers, some users may drop from the training process so that the users only wish to securely compute the sum of the surviving users from the server's reply messages. We show that if any user may drop, to cope with such uncertainty, each user needs to hold a key of $K$ bits per sum bit (so essentially storing all keys) while other optimal rates remain the same as those with no user dropouts.

\section{Problem Statement}
\subsection{Secure Aggregation with an Oblivious Server} \label{sec:oblivious}
Consider $K \geq 2$ users and User $k \in \{1,2,\cdots, K\} \triangleq [K]$ holds input $W_k$ and key $Z_k$. The inputs $\left(W_k\right)_{k\in[K]}$ are independent and are independent of $\left(Z_k\right)_{k\in[K]}$. Each $W_k$ is an $L \times 1$ column vector where the $L$ elements are i.i.d. uniform\footnote{To facilitate the presentation of the Shannon theoretic model, we assume statistical inputs. Specifically, we assume that the inputs are uniform, which is critical for the converse proofs while our achievable scheme works for arbitrarily distributed inputs. One may use an equivalent deterministic input model where the inputs can be arbitrary sequences (see e.g., \cite{Zhao_Sun_Expand}).} over the finite field $\mathbb{F}_q$. 
\begin{eqnarray}
   && H\left(\left(W_k\right)_{k\in[K]},
    \left(Z_k\right)_{k\in[K]}\right)=
    \sum_{k\in[K]} H\left(W_k \right) +
    H\left(\left(Z_k\right)_{k\in[K]} \right), \label{ind} \\
   && H(W_k) = L ~(\mbox{in $q$-ary units}), ~\forall k \in [K]. \label{h2}
\end{eqnarray}
Each $Z_k$ is comprised of $L_Z$ symbols from $\mathbb{F}_q$. $\left(Z_k\right)_{k\in[K]}$ are a function of a source key variable $Z_\Sigma$, which is comprised of $L_{Z_{\Sigma}}$ symbols from $\mathbb{F}_q$.
\begin{eqnarray}
	H\left(\left(Z_k\right)_{k\in[K]} \big| Z_\Sigma\right) = 0.\label{zsum}
\end{eqnarray}
Consider a server.
The communication protocol includes one message from each user to the server and one message from the server to each user. Specifically, in the first phase, User $k$ sends a message $X_k$, $k \in [K]$ to the server. The message $X_k$ is a function of $W_k, Z_k$ and consists of $L_X$ symbols from $\mathbb{F}_q$.
\begin{eqnarray}
    H\left(X_k | W_k, Z_k\right) = 0, \forall k \in [K].\label{messagex}
\end{eqnarray}

In the second phase, the server sends a message $Y_k, k \in [K]$ to User $k$. The message $Y_k$ is a function of $\left(X_k\right)_{k\in[K]}$ (what the server just received) and consists of $L_Y$ symbols from $\mathbb{F}_q$.
\begin{eqnarray}
    H\left(Y_k \big| \left(X_u\right)_{u\in[K]} \right) = 0, \forall k \in [K].\label{messagey}
\end{eqnarray}

After receiving the message from the server, each user must be able to recover the desired sum $\sum_{k \in [K]} W_k$ with no error, combined with its own input and key.
\begin{eqnarray}
    \mbox{[Correctness]}~~~H\left(\sum_{u\in[K]} W_u \Bigg| Y_k, W_k, Z_k \right) = 0, \forall k \in [K].\label{corr}
\end{eqnarray}
Security refers to the constraint that the server cannot infer any information about $\left(W_k\right)_{k\in[K]}$ and each user cannot obtain any information about $\left(W_k\right)_{k\in[K]}$ beyond that contained in the desired sum and known by itself. That is, the following security constraint must be satisfied.
\begin{eqnarray}
    \mbox{[Server Security]} && I\left(\left(W_k\right)_{k\in[K]}; \left(X_k\right)_{k\in[K]} 
    \right) = 0.\label{security_server}\\
        \mbox{[User Security]} && I\left(\left(W_u\right)_{u\in[K]}; Y_k \Bigg| \sum_{u\in [K]} W_u, W_k, Z_k \right) = 0, \forall k \in [K]. \label{security_user}
\end{eqnarray}
The communication {\em rate} $R_X$ ($R_Y$) characterizes how many symbols message $X_k$ ($Y_k$) contains per sum (input) symbol, and is defined as follows.
\begin{eqnarray}
	R_X \triangleq \frac{L_X}{L}, ~R_Y \triangleq \frac{L_Y}{L}. \label{rate:R}
\end{eqnarray}
The individual (total) key {\em rate} $R_{Z}$ ($R_{Z_{\Sigma}}$) characterizes how many symbols key $Z_k$ (source key $Z_\Sigma$) contains per sum symbol, and is defined as follows.
\begin{eqnarray}
    R_{Z} \triangleq \frac {L_{Z}}{L}, ~R_{Z_{\Sigma}} \triangleq \frac {L_{Z_{\Sigma}}}{L}. \label{rate:RZ}
\end{eqnarray}
A rate tuple $(R_X, R_Y, R_{Z},R_{Z_{\Sigma}})$ is said to be achievable if there exists a secure aggregation scheme (i.e., a design of keys $Z_k, Z_{\Sigma}$ and messages $X_k, Y_k$), for which the correctness constraint (\ref{corr}) and the security constraint (\ref{security_server}), (\ref{security_user}) are satisfied, and the first phase and second phase message rates, the individual key rate, and the total key rate are no greater than $R_X, R_Y, R_{Z}$, and $R_{Z_{\Sigma}}$, respectively. The closure of the set of all achievable rate tuples is called the optimal rate region, 
denoted as $\mathcal{R}^{*}$.

\subsection{Secure Aggregation with an Oblivious Server and User Dropouts}\label{sec:oblivious_drop}
We generalize the model in the previous section to include user dropouts. The assumption on the inputs and keys is the same (refer to (\ref{ind}), (\ref{h2}), (\ref{zsum})). The first phase (i.e., $X_k$ messages) is also identical as at the beginning all $K$ users are present (refer to (\ref{messagex})). At the end of the first phase, the server may not receive all $X_k$ messages such that the missing users are viewed as dropped (from the training process) and each user now desires the sum of the inputs from the surviving users. The set of surviving users is denoted as $\mathcal{U}$, where 
$\mathcal{U}$ is an arbitrary subset of $[K]$ 
so that 
any set of users might drop. 

We proceed to the second phase, where the server sends a message $Y_k^{\mathcal{U}}, k \in \mathcal{U}$ of $L_Y$ symbols from $\mathbb{F}_q$ to surviving User $k$. 
\begin{eqnarray}
    H\left(Y_k^{\mathcal{U}} \big| \left(X_u\right)_{u\in\mathcal{U}} \right) = 0, \forall k \in \mathcal{U} \label{messagey_drop}
\end{eqnarray}
where the server informs each user the surviving user set $\mathcal{U}$ (whose communication cost is negligible and omitted as it does not depend on the input/block size $L$), indicated by the superscript of $Y_k$.
The correctness constraint specifies the decodability of the sum of surviving inputs.
\begin{eqnarray}
    \mbox{[Correctness]}~~~H\left(\sum_{u\in \mathcal{U}} W_u \Bigg| Y_k^\mathcal{U}, W_k, Z_k \right) = 0, \forall k \in \mathcal{U}.\label{corr_drop}
\end{eqnarray}
Regarding security, the server security constraint is the same as (\ref{security_server}) and the user security constraint is modified accordingly as follows.
\begin{eqnarray}
        \mbox{[User Security]} && I\left(\left(W_u\right)_{u\in [K]}; Y_k^{\mathcal{U}} \Bigg| \sum_{u\in \mathcal{U}} W_u, W_k, Z_k \right) = 0, \forall k \in \mathcal{U}. \label{security_user_drop}
\end{eqnarray}

The definition of the rates $R_X, R_Y, R_Z, R_{Z_\Sigma}$, the achievable rate tuple, and the optimal rate region $\mathcal{R}^*$ is standard Shannon theoretic and is identical to that above.

\section{Results}
In this section, we summarize our main results 
along with key observations. 
The optimal rate region of secure aggregation with an oblivious server is characterized in Theorem \ref{thm:oblivious}, presented below.
\begin{theorem}\label{thm:oblivious}
For the secure aggregation with an oblivious server problem defined in Section \ref{sec:oblivious}, the optimal rate region is
\begin{eqnarray}
    \mathcal{R}^* =
    \left\{ \left(R_X, R_Y, R_{Z}, R_{Z_{\Sigma}}\right) : R_X \geq 1, R_Y \geq 1, R_{Z} \geq 2, R_{Z_{\Sigma}} \geq K\right\}.
\end{eqnarray}
\end{theorem}

An intuitive explanation of Theorem \ref{thm:oblivious} may be seen as follows. Suppose we wish to securely compute $1$ symbol of the sum. Each user needs to send at least $1$ symbol to the server because the message must carry its own input ($1$ symbol) which is independent of all other inputs and appears in the desired sum (thus $R_X \geq 1$). Each server needs to send at least $1$ symbol to each user because the message must carry the sum of all other inputs ($1$ symbol), for each user to correctly decode the sum (thus $R_Y \geq 1$). Note that these two message rate bounds do not need the security constraint as the arguments are based solely on the correctness constraint (i.e., missing information for sum computation).
Next, each user needs to hold a key of at least $2$ symbols because we need to protect the ($1$ symbol) message from the user to the server and also the ($1$ symbol) message from the server to the user. Furthermore, these two key symbols must be independent as otherwise the server will learn something about the inputs (thus $R_Z \geq 2$, it turns out that this inequality is the most challenging to prove). Finally, all users must hold at least $K$ key symbols because we need to protect all $K$ message symbols from the users to the server (reminiscence of Shannon's information theoretic security result on point-to-point secure communication \cite{shannon1949}). The detailed proof of Theorem \ref{thm:oblivious} is presented in Section \ref{proof:oblivious}. 

\vspace{0.1in}
The optimal rate region of secure aggregation with an oblivious server and user dropouts is characterized in Theorem \ref{thm:oblivious_drop}, presented below.

\begin{theorem}\label{thm:oblivious_drop}
For the secure aggregation with an oblivious server and user dropouts problem defined in Section \ref{sec:oblivious_drop}, the optimal rate region is
\begin{eqnarray}
    \mathcal{R}^* =
    \left\{ \left(R_X, R_Y, R_{Z}, R_{Z_{\Sigma}}\right) : R_X \geq 1, R_Y \geq 1, R_{Z} \geq K, R_{Z_{\Sigma}} \geq K\right\}.
\end{eqnarray}
\end{theorem}

Compared to Theorem \ref{thm:oblivious}, only the individual key rate result changes in Theorem \ref{thm:oblivious_drop} which may be seen intuitively as follows. We use the notation $\mathcal{A} \setminus \mathcal{B}$ to denote the set of elements that belong to $\mathcal{A}$ but not $\mathcal{B}$.
As any user might drop, in the second phase User $1$ will need to be able to decode $\sum_{k \in [K] \setminus \{u\}} W_k$ for any $u \neq 1$ (when User $u$ drops) and $\sum_{k=1}^K W_k$ (when no user drops), the collection of which is invertible to all $K$ inputs (refer to Lemma \ref{lemma:inv}). Then we may argue that each user needs to prepare $K$ key symbols for all these cases (and keys cannot be reused as that will introduce undesired key correlation and violate server security). The detailed proof of Theorem \ref{thm:oblivious_drop} is presented in Section \ref{proof:oblivious_drop}. 


\section{Proof of Theorem \ref{thm:oblivious}}\label{proof:oblivious}
\subsection{Converse}
{\em Proof of $R_X \geq 1$:} Let us start with a lemma which will easily lead to the desired bound on $R_X$ and will also be useful in the proof of other rate bounds.

We show that each $X_k$ must contribute all information about $W_k$ and thus contain at least $L$ symbols. 
The proof generalizes that of Lemma 1 in \cite{Zhao_Sun_Sum}.
\begin{lemma}\label{lemma:xu}
For any $u \in [K]$, we have
\begin{eqnarray}
    && H\left(X_u|Z_u, (W_k,Z_k)_{k\in[K]\setminus \{u\}}\right) = L, \label{eq:xu} \\
    && H(X_1 | Z_1) \geq L, \label{eq:x1z1} \\
    && H\left( \left(X_k\right)_{k \in [K]} \right) \geq KL. \label{eq:xk}
\end{eqnarray}
\end{lemma}

\proof First, consider (\ref{eq:xu}). As $K \geq 2$, there exists $v$ such that $v \neq u$ and $v \in [K]$.
\begin{eqnarray}
    && H\left(X_u|Z_u, (W_k,Z_k)_{k\in[K]\setminus \{u\}}\right)\notag\\
    &\overset{(\ref{messagex})}{=}& I\left(X_u;  W_u \Big| Z_u, (W_k,Z_k)_{k\in[K]\setminus \{u\}}\right)  \label{eq:e0} \\
    &=& H\left( W_u \Big| Z_u, (W_k,Z_k)_{k\in[K]\setminus \{u\}}\right) - H\left( W_u \Big| X_u, Z_u, (W_k,Z_k)_{k\in[K]\setminus \{u\}}\right) \\
    &\overset{(\ref{ind})(\ref{messagex})(\ref{messagey})(\ref{corr})}{=}& H\left(W_u\right) - 
    H\left(W_u \Bigg| (X_k)_{k\in[K]}, Z_u, Y_v, W_v, Z_v, \sum_{k\in[K]} W_k, (W_k, Z_k)_{k\in[K]\setminus \{u\}} \right)
    \label{pf_lemma1_1}\\
    &\overset{(\ref{h2})}{=}& L \label{eq:e1} 
\end{eqnarray}
where (\ref{eq:e0}) is due to the fact that $X_u$ is determined by $W_u, Z_u$ (see (\ref{messagex})). For (\ref{pf_lemma1_1}), the first term follows from the fact that input $W_u$ is independent of other inputs and all keys (see (\ref{ind})); the second term follows from the fact that $(X_k)_{k\in[K]\setminus \{u\}}$ is determined by $(W_k,Z_k)_{k\in[K]\setminus \{u\}}$ (see (\ref{messagex})), $Y_v$ is determined by $(X_k)_{k\in[K]}$ (see (\ref{messagey})), and the sum $\sum_{k\in[K]} W_k$ can be decoded by User $v$ from $Y_v, W_v, Z_v$ (see (\ref{corr})). In (\ref{eq:e1}), we use the property that $W_u$ has $L$ uniform symbols (see (\ref{h2})) and $W_u$ can be recovered from the sum $\sum_{k\in[K]} W_k$ and all other inputs $(W_k)_{k\in[K] \setminus \{u\}}$ (such that the second term of (\ref{pf_lemma1_1}) is zero).

Second, (\ref{eq:x1z1}) follows from setting $u=1$ in (\ref{eq:xu}) and the property that dropping conditioning cannot decrease entropy. 

Finally, we show that (\ref{eq:xk}) is a simple consequence of (\ref{eq:xu}).
\begin{eqnarray}
H\left( \left(X_k\right)_{k \in [K]} \right) &\geq& \sum_{u \in [K]} H\left( X_u \Big| \left(X_k\right)_{k \in [K] \setminus \{u\}} \right) \\
&\geq& \sum_{u \in [K]} H\left( X_u \Big| \left(X_k, W_k, Z_k \right)_{k \in [K] \setminus \{u\}} \right) \\
&\overset{(\ref{messagex})}{=}& \sum_{u \in [K]} H\left( X_u \Big| \left(W_k, Z_k \right)_{k \in [K] \setminus \{u\}} \right) \\ 
&\overset{(\ref{eq:xu})}{=}& KL.
\end{eqnarray}
\hfill\QED

Consider $R_X$ now.
For any $u \in [K]$, we have
\begin{eqnarray}
    L_X \geq H\left(X_u\right) \overset{(\ref{eq:xu})}{\geq} L 
    ~~\Rightarrow~~ R_X \overset{(\ref{rate:R})}{=} L_X/L \geq 1.
\end{eqnarray}

{\em Proof of $R_Y \geq 1$:} This proof on $R_Y$ is a minor variation of that on $R_X$ presented above.

Consider any $u \in [K]$.
\begin{eqnarray}
    && L_Y \geq H(Y_u) \geq H\left(Y_u| W_u, (Z_k)_{k\in[K]}\right) \\
    &\geq& I\left(Y_u;  \sum_{k\in[K]} W_k \Bigg| W_u, (Z_k)_{k\in[K]}\right)\\
    &=& H\left( \sum_{k\in[K]} W_k \Bigg| W_u, (Z_k)_{k\in[K]}\right) - H\left( \sum_{k\in[K]} W_k \Big| Y_u, W_u,  (Z_k)_{k\in[K]}\right) \\
    &\overset{(\ref{ind})(\ref{corr})}{=}& H\left(\sum_{k\in[K] \setminus \{u\}} W_k \right) - 
    H\left(\sum_{k\in[K]} W_k \Bigg| Y_u, W_u, (Z_k)_{k\in[K]}, \sum_{k\in[K]} W_k \right)
    \\
    &\overset{(\ref{ind})(\ref{h2})}{=}& L - 0 = L\label{eq:e2} 
    \\
    &\Rightarrow& R_Y ~\overset{(\ref{rate:R})}{=}~ L_Y/L ~\geq~ 1
\end{eqnarray}
where (\ref{eq:e2}) follows from the observation that the sum of any subset of $W_k$ is uniform, as $(W_k)_{k\in[K]}$ are i.i.d. uniform.

\vspace{0.1in}
{\em Proof of $R_{Z} \geq 2$:} This proof can be viewed as a generalization of Shannon's result on secure communication to include computation tasks \cite{shannon1949}.
\begin{eqnarray}
&& L_Z \geq H(Z_1) \geq H(Z_1 | X_1, Y_1) \\
&=& H\left(Z_1, W_1, \sum_{k\in[K]} W_k \Bigg| X_1, Y_1\right) - H\left(W_1, \sum_{k\in[K]} W_k \Bigg| X_1, Y_1, Z_1 \right) \\
&\geq& H\left(W_1, \sum_{k\in[K]} W_k \Bigg| X_1, Y_1\right) - H\left(W_1 | X_1, Y_1, Z_1 \right) 
- \underbrace{ H\left(\sum_{k\in[K]} W_k \Bigg| X_1, Y_1, Z_1, W_1 \right) }_{ \overset{(\ref{corr})}{=} 0} \\
&\geq& H\left(W_1, \sum_{k\in[K]} W_k \Bigg| (X_k)_{k\in[K]}, Y_1\right) - H\left(W_1 | X_1, Z_1 \right) \\ 
&\overset{(\ref{messagey})}{=}& H\left(W_1, \sum_{k\in[K]} W_k \Bigg| (X_k)_{k\in[K]} \right) - H\left(W_1 | Z_1 \right) + I\left(W_1; X_1 | Z_1 \right) \\
&\overset{(\ref{security_server})(\ref{ind})(\ref{messagex})}{=}& H\left(W_1, \sum_{k\in[K]} W_k \right) - H(W_1) + H(X_1 | Z_1) \label{eq:e3} \\
&\overset{(\ref{ind})(\ref{h2})(\ref{eq:x1z1})}{\geq}& 2L - L + L = 2L\\
&\Rightarrow& R_Z ~\overset{(\ref{rate:RZ})}{=}~ L_Z/L \geq 2
\end{eqnarray}
where the first term of (\ref{eq:e3}) follows from the server security constraint (\ref{security_server}), i.e., $(X_k)_{k\in[K]}$ is independent of $(W_k)_{k\in[K]}$ (thus also any function of $(W_k)_{k\in[K]}$).

{\em Proof of $R_{Z_{\Sigma}} \geq K$:}
\begin{eqnarray}
&& L_{Z_\Sigma} \geq H(Z_\Sigma) \overset{(\ref{zsum})}{=} H\left(Z_\Sigma, (Z_k)_{k\in[K]} \right)\\
&\geq& H\left( (Z_k)_{k\in[K]} \Big| (W_k)_{k\in[K]} \right) ~\geq~ I\left( (X_k)_{k\in[K]} ; (Z_k)_{k\in[K]} \Big| (W_k)_{k\in[K]} \right) \\
&\overset{(\ref{messagex})}{=}& H\left( (X_k)_{k\in[K]} \Big| (W_k)_{k\in[K]} \right) \\
&\overset{(\ref{security_server})}{=}& H\left( (X_k)_{k\in[K]} \right) \\
&\overset{(\ref{eq:xk})}{\geq}& KL \\
&\Rightarrow& R_{Z_\Sigma} \overset{(\ref{rate:RZ})}{=} L_{Z_\Sigma}/L ~\geq~ K.
\end{eqnarray}

{\it Remark:} Interestingly, we may notice that the above converse proof does not use the user security constraint (\ref{security_user}), so user security is obtained for free, i.e., even if user security is removed, we cannot achieve a better rate.

\subsection{Achievability}
The achievable scheme is plotted in Fig.~\ref{fig:oblivious} and here is the proof for completeness.

Consider $K$ i.i.d. uniform variables over $\mathbb{F}_q$, $N_1,\cdots,N_{K}$. Set the keys and messages as 
 \begin{eqnarray}
 Z_\Sigma &=& \left( N_1, N_2, \cdots, N_K \right), \\
 Z_k &=& \left( N_k, \sum_{u\in[K]}N_u \right), \label{eq:zk} \\
 X_k &=& W_k + N_k, \label{eq:xk_ach} \\
 Y_k &=& \sum_{u\in[K]} X_k = \sum_{u\in[K]} W_u + \sum_{u\in[K]} N_u, \forall k \in [K] \label{eq:yk}
 \end{eqnarray}
so  that $L=1, L_X=1, L_Y=1, L_Z = 2, L_{Z_\Sigma} = K$ and the rate achieved is $R_X = 1, R_Y = 1, R_Z = 2, R_{Z_\Sigma} = K$, as desired. 

Correctness is easily seen, as $\sum_{u\in[K]} W_u = Y_k - \sum_{u\in[K]} N_u$. Finally, we verify that the security constraint is satisfied. For server security (\ref{security_server}),
\begin{eqnarray}
 I\left(\left(W_k\right)_{k\in[K]}; \left(X_k\right)_{k\in[K]} \right) &\overset{(\ref{eq:xk_ach})}{=}&  I\left(\left(W_k\right)_{k\in[K]}; \left(W_k + N_k\right)_{k\in[K]} \right) \label{eq:a0}\\
 &=& H \left( \left(W_k + N_k\right)_{k\in[K]} \right) - H \left(\left(W_k + N_k\right)_{k\in[K]} | \left(W_k\right)_{k\in[K]} \right) \\
 &\overset{(\ref{ind})}{\leq}& KL - H \left( \left(N_k\right)_{k\in[K]} \right) \label{eq:a1} \\
 &=& KL - KL \label{eq:a2} \\
 &=& 0
\end{eqnarray}
where the first term of (\ref{eq:a1}) is due to the fact that uniform distribution maximizes entropy and the second term follows from the independence of $(N_k)_{k\in[K]}$ and $(W_k)_{k\in[K]}$. The second term of (\ref{eq:a2}) uses the uniformity of $(N_k)_{k\in[K]}$. Note that as mutual information is non-negative, the above derivation shows that the term on the LHS of (\ref{eq:a0}) must be zero.

For user security (\ref{security_user}),
\begin{eqnarray}
&& I\left(\left(W_u\right)_{u\in[K]}; Y_k \Bigg| \sum_{u\in [K]} W_u, W_k, Z_k \right) \notag \\
&\overset{(\ref{eq:yk})(\ref{eq:zk})}{=}& I\left(\left(W_u\right)_{u\in[K]}; \sum_{u\in[K]} \left(W_u + N_u \right) \Bigg| \sum_{u\in [K]} W_u, W_k, N_k, \sum_{u\in[K]} N_u \right) \\
&=& 0 
\end{eqnarray}
where the last step follows from the observation that the conditioning terms determine $Y_k$.

{\it Remark:} Note that the above achievability proof does not use the uniformity of $(W_k)_{k\in[K]}$ so that it works for any distribution of $(W_k)_{k\in[K]}$.

\section{Proof of Theorem \ref{thm:oblivious_drop}}\label{proof:oblivious_drop}
\subsection{Converse}
As the converse bounds on $R_X, R_Y, R_{Z_\Sigma}$ do not change when compared to the no user dropout case (refer to Theorem \ref{thm:oblivious}), the same proof works when we set $\mathcal{U} = [K]$ (i.e., the constraints now reduce to the same as those in Section \ref{sec:oblivious}) and thus we only need to prove $R_Z \geq K$. To this end, we first prove a lemma on the invertibility between a vector of subset sums of the inputs and all inputs. Define 
\begin{eqnarray}
{\vec W} = \left(\sum_{k\in[K]} W_k, \sum_{k\in[K]\setminus\{K\}} W_k, \sum_{k\in[K]\setminus\{K-1\}} W_k, \cdots, \sum_{k\in[K]\setminus\{3\}} W_k \right).
\end{eqnarray}

\begin{lemma}\label{lemma:inv}
$(W_1, {\vec W})$ is invertible to $(W_k)_{k\in[K]}$.
\end{lemma}

{\proof} Obviously, from $(W_k)_{k\in[K]}$ we may recover $(W_1, {\vec W})$ and we now prove the reverse direction. As we know $W_1$ and $\sum_{k\in[K]} W_k$, consider the remaining $K-2$ terms and we have
\begin{eqnarray}
&& \sum_{k\in[K]} W_k - \sum_{k\in[K]\setminus\{K\}} W_k = W_K, \\
&& \sum_{k\in[K]} W_k - \sum_{k\in[K]\setminus\{K-1\}} W_k = W_{K-1}, \\
&& ~~~~~~~~~~~~\vdots\\
&& \sum_{k\in[K]} W_k - \sum_{k\in[K]\setminus\{3\}} W_k = W_{3}.
\end{eqnarray}
Combining with $W_1$ and $\sum_{k\in[K]} W_k$, we may recover $(W_k)_{k\in[K]}$ and the proof is complete.

\hfill\QED

The remaining steps of the converse proof are based on a combination of Lemma \ref{lemma:inv} and the $R_Z$ proof in Theorem \ref{thm:oblivious}. Corresponding to the $K-1$ terms in ${\vec W}$, consider the following choices of surviving user set, $\mathcal{U}_1 = [K]$, $\mathcal{U}_2 = [K] \setminus \{K\}$, $\mathcal{U}_3 = [K] \setminus \{K-1\}$, $\cdots$, $\mathcal{U}_{K-1} = [K] \setminus \{3\}$.

\begin{eqnarray}
&& L_Z \geq H(Z_1) \geq H\left(Z_1 \Big| X_1, \left(Y_1^{\mathcal{U}_v} \right)_{v \in [K-1]} \right) \\
&=& H\left(Z_1, W_1, {\vec W} \Big| X_1, \left(Y_1^{\mathcal{U}_v} \right)_{v \in [K-1]} \right) - H\left(W_1, {\vec W} \Big| X_1, \left(Y_1^{\mathcal{U}_v} \right)_{v \in [K-1]}, Z_1 \right) \\
&\geq& H\left(W_1, {\vec W} \Big| X_1, \left(Y_1^{\mathcal{U}_v} \right)_{v \in [K-1]} \right) - H\left(W_1 \Big| X_1,  \left(Y_1^{\mathcal{U}_v} \right)_{v \in [K-1]}, Z_1 \right) \notag\\
&&~ 
- \underbrace{ H\left( {\vec W} \Big| X_1,  \left(Y_1^{\mathcal{U}_v} \right)_{v \in [K-1]}, Z_1, W_1 \right) }_{ \overset{(\ref{corr_drop})}{=} 0} \label{eq:c1} \\
&\geq& H\left( \left(W_k\right)_{k\in[K]} \Big| (X_k)_{k\in[K]},  \left(Y_1^{\mathcal{U}_v} \right)_{v \in [K-1]} \right) - H\left(W_1 | X_1, Z_1 \right) \label{eq:c2}\\ 
&\overset{(\ref{messagey_drop})}{=}& H\left( \left(W_k\right)_{k\in[K]} \Big| (X_k)_{k\in[K]} \right) - H\left(W_1 | Z_1 \right) + I\left(W_1; X_1 | Z_1 \right) \\
&\overset{(\ref{security_server})(\ref{ind})(\ref{messagex})}{=}& H\left( \left(W_k\right)_{k\in[K]} \right) - H(W_1) + H(X_1 | Z_1) \\
&\overset{(\ref{ind})(\ref{h2})(\ref{eq:x1z1})}{\geq}& KL - L + L = KL\\
&\Rightarrow& R_Z ~\overset{(\ref{rate:RZ})}{=}~ L_Z/L \geq K
\end{eqnarray}
where the last term of (\ref{eq:c1}) follows because the choice of surviving user sets matches ${\vec W}$ and the first term of (\ref{eq:c2}) is due to Lemma \ref{lemma:inv}.

\subsection{Achievability}
Consider $K$ i.i.d. uniform variables over $\mathbb{F}_q$, $N_1,\cdots,N_{K}$. Set the keys and messages as 
 \begin{eqnarray}
 Z_\Sigma &=& Z_k ~=~ \left( N_1, N_2, \cdots, N_K \right),  \forall k \in [K], \label{eq:zk_drop}\\
 X_k &=& W_k + N_k, \forall k \in [K], \label{eq:xk_ach_drop} \\
 Y_k^\mathcal{U} &=& \sum_{u\in\mathcal{U}} X_u = \sum_{u\in\mathcal{U}} W_u + \sum_{u\in\mathcal{U}} N_u, \forall k \in \mathcal{U}, \forall \mathcal{U} \subset [K] \label{eq:yk_drop}
 \end{eqnarray}
so  that $L=1, L_X=1, L_Y=1, L_Z = K, L_{Z_\Sigma} = K$ and the rate achieved is $R_X = 1, R_Y = 1, R_Z = K, R_{Z_\Sigma} = K$, as desired. 

Correctness holds because $\sum_{u\in\mathcal{U}} W_u = Y_k^\mathcal{U} - \sum_{u\in\mathcal{U}} N_u$ and $(N_k)_{k\in[K]}$ is known by every user. Finally, we verify security. Server security is the same as that of Theorem \ref{thm:oblivious}. User security (\ref{security_user_drop}) is similarly proved as follows.
\begin{eqnarray}
&& I\left(\left(W_u\right)_{u\in[K]}; Y_k^\mathcal{U} \Bigg| \sum_{u\in \mathcal{U}} W_u, W_k, Z_k \right) \\
&\overset{(\ref{eq:yk_drop})(\ref{eq:zk_drop})}{=}& I\left(\left(W_u\right)_{u\in\mathcal{U}}; \sum_{u \in\mathcal{U}} \left(W_u + N_u \right) \Bigg| \sum_{u\in \mathcal{U}} W_u, W_k, (N_u)_{u\in[K]} \right) \\
&=& 0.
\end{eqnarray}

\section{Discussion}
In this work, we have characterized the optimal rate region of secure aggregation with an oblivious server, and with or without user dropouts. The model considered is elementary and below we discuss a few natural generalizations.

{\bf Colluding Users:} Consider the situation where we allow the server to collude with some subset of users to try to infer additional information about remaining users. When no users might drop (refer to Section \ref{sec:oblivious}), it is straightforward to check that the achievable scheme from the proof of Theorem \ref{thm:oblivious} guarantees the security of the non-colluding users so that user collusion does not hurt the rate region. When users might drop (refer to Section \ref{sec:oblivious_drop}), then even if the server colludes with only one user, they can still recover all $K$ inputs because we may consider the surviving user sets in the converse proof of Theorem \ref{thm:oblivious_drop} (refer to the paragraph below the proof of Lemma \ref{lemma:inv}) and the decoded subset sums are invertible to all $K$ inputs (refer to Lemma \ref{lemma:inv}), making user collusion a trivial setting as nothing can be hidden.

{\bf Broadcast Server Message:} In the second phase when the server sends messages to the users, we assume the communication link is unicast, i.e., the server may send a different message to each user. An alternative choice is a broadcast channel model where the server sends the same message to each user, which is viewed as more efficient in many practical scenarios. Interestingly, we note that our achievable scheme in the proof of both Theorem \ref{thm:oblivious} and Theorem \ref{thm:oblivious_drop} sends the same message from the server to all users and with minor variations both converse proofs work under the broadcast channel model. Therefore,  the optimal rate region remains the same if the server to user communication link is a broadcast channel.

{\bf Randomness at Server:} We have assumed that the server is not equipped with any key and what if the server holds some key variables that are arbitrarily correlated with the users' keys? With some straightforward modification (e.g., by adding conditioning on the server's key), one may find that the converse proofs go through so that the same converse bounds hold and we may apply the same achievable scheme. Thus the optimal rate region remains the same, which shall make intuitive sense because we wish to be secure against the server so that giving the server more power (key variables) should not help. 

{\bf User Dropout Pattern:} We allow arbitrary users to drop in the problem statement (refer to Section \ref{sec:oblivious_drop}). Now what if we set beforehand the possible user dropout patterns? Following the converse proof of Theorem \ref{thm:oblivious_drop} (especially Lemma \ref{lemma:inv}, which only uses single user dropout), we can see that it will be related to the collection of inputs that can be possibly decoded by one user for all possible user dropout patterns. As long as the user dropout pattern is rich to the extent that all $K$ inputs are recoverable, then the same individual key rate bound holds. Otherwise, the key rate bound will depend on the rank of the recoverable inputs.
Communication rates and total key rate will not be influenced.

Perhaps the most interesting aspect of this work is the research line of understanding the information theoretic limits of the paradigm of adding oblivious helpers in computation tasks such as secure aggregation. Oblivious computing helpers might be useful in enhancing security, reducing communication, storage, or randomness cost. As potential future topics, it is worthwhile to study the information theoretic benefits and limitations of oblivious helpers in more variants of secure aggregation such as user selection \cite{selection}, weaker security criteria \cite{Zhou_Zhao_Sun, li2022arithmetic}, and groupwise keys \cite{Zhao_Sun_Aggregate, WSJC_Groupwise}, and more generally other computation tasks such as federated submodel learning \cite{vithana2023private, wang2023fully}.

\bibliography{Thesis}

\begin{thebibliography}{10}
\providecommand{\url}[1]{#1}
\csname url@samestyle\endcsname
\providecommand{\newblock}{\relax}
\providecommand{\bibinfo}[2]{#2}
\providecommand{\BIBentrySTDinterwordspacing}{\spaceskip=0pt\relax}
\providecommand{\BIBentryALTinterwordstretchfactor}{4}
\providecommand{\BIBentryALTinterwordspacing}{\spaceskip=\fontdimen2\font plus
\BIBentryALTinterwordstretchfactor\fontdimen3\font minus
  \fontdimen4\font\relax}
\providecommand{\BIBforeignlanguage}[2]{{%
\expandafter\ifx\csname l@#1\endcsname\relax
\typeout{** WARNING: IEEEtran.bst: No hyphenation pattern has been}%
\typeout{** loaded for the language `#1'. Using the pattern for}%
\typeout{** the default language instead.}%
\else
\language=\csname l@#1\endcsname
\fi
#2}}
\providecommand{\BIBdecl}{\relax}
\BIBdecl

\bibitem{aggregation}
K.~Bonawitz, V.~Ivanov, B.~Kreuter, A.~Marcedone, H.~B. McMahan, S.~Patel,
  D.~Ramage, A.~Segal, and K.~Seth, ``{Practical Secure Aggregation for
  Privacy-Preserving Machine Learning},'' in \emph{Proceedings of the 2017 ACM
  SIGSAC Conference on Computer and Communications Security}, 2017, pp.
  1175--1191.

\bibitem{Zhao_Sun_Aggregate}
Y.~Zhao and H.~Sun, ``{Information Theoretic Secure Aggregation With User
  Dropouts},'' \emph{IEEE Transactions on Information Theory}, vol.~68, no.~11,
  pp. 7471--7484, 2022.

\bibitem{aggregation_light}
J.~So, C.~J. Nolet, C.-S. Yang, S.~Li, Q.~Yu, R.~E~Ali, B.~Guler, and
  S.~Avestimehr, ``{LightSecAgg: a Lightweight and Versatile Design for Secure
  Aggregation in Federated Learning},'' \emph{Proceedings of Machine Learning
  and Systems}, vol.~4, pp. 694--720, 2022.

\bibitem{aggregation_swift}
T.~Jahani-Nezhad, M.~A. Maddah-Ali, S.~Li, and G.~Caire, ``{SwiftAgg+:
  Achieving Asymptotically Optimal Communication Loads in Secure Aggregation
  for Federated Learning},'' \emph{IEEE Journal on Selected Areas in
  Communications}, vol.~41, no.~4, pp. 977--989, 2023.

\bibitem{WSJC_Groupwise}
K.~Wan, H.~Sun, M.~Ji, and G.~Caire, ``{Information Theoretic Secure
  Aggregation with Uncoded Groupwise Keys},'' \emph{arXiv preprint
  arXiv:2204.11364}, 2022.

\bibitem{Zhao_Sun_Sum}
Y.~Zhao and H.~Sun, ``{Secure Summation: Capacity Region, Groupwise Key, and
  Feasibility},'' \emph{arXiv preprint arXiv:2205.08458}, 2022.

\bibitem{zhao2023secure}
J.~C. Zhao, A.~Sharma, A.~R. Elkordy, Y.~H. Ezzeldin, S.~Avestimehr, and
  S.~Bagchi, ``{Secure Aggregation in Federated Learning is not Private:
  Leaking User Data at Large Scale through Model Modification},'' \emph{arXiv
  preprint arXiv:2303.12233}, 2023.

\bibitem{BGW}
M.~Ben-Or, S.~Goldwasser, and A.~Wigderson, ``{Completeness Theorems for
  Non-Cryptographic Fault-Tolerant Distributed Computation},'' in
  \emph{Proceedings of the twentieth annual ACM symposium on Theory of
  computing}.\hskip 1em plus 0.5em minus 0.4em\relax ACM, 1988, pp. 1--10.

\bibitem{CCD}
D.~Chaum, C.~Cr{\'e}peau, and I.~Damgard, ``{Multiparty Unconditionally Secure
  Protocols},'' in \emph{Proceedings of the twentieth annual ACM symposium on
  Theory of computing}.\hskip 1em plus 0.5em minus 0.4em\relax ACM, 1988, pp.
  11--19.

\bibitem{cramer_damgard_nielsen_2015}
R.~Cramer, I.~B. Damgard, and J.~B. Nielsen, \emph{Secure Multiparty
  Computation and Secret Sharing}.\hskip 1em plus 0.5em minus 0.4em\relax
  Cambridge University Press, 2015.

\bibitem{hirt2000player}
M.~Hirt and U.~Maurer, ``{Player Simulation and General Adversary Structures in
  Perfect Multiparty Computation},'' \emph{Journal of cryptology}, vol.~13,
  no.~1, pp. 31--60, 2000.

\bibitem{cramer2005share}
R.~Cramer, I.~Damg{\aa}rd, and Y.~Ishai, ``{Share Conversion, Pseudorandom
  Secret-Sharing and Applications to Secure Computation},'' in \emph{Theory of
  Cryptography: Second Theory of Cryptography Conference, TCC 2005, Cambridge,
  MA, USA, February 10-12, 2005. Proceedings 2}.\hskip 1em plus 0.5em minus
  0.4em\relax Springer, 2005, pp. 342--362.

\bibitem{garay2017price}
J.~Garay, Y.~Ishai, R.~Ostrovsky, and V.~Zikas, ``{The Price of Low
  Communication in Secure Multi-Party Computation},'' in \emph{Advances in
  Cryptology--CRYPTO 2017: 37th Annual International Cryptology Conference,
  Santa Barbara, CA, USA, August 20--24, 2017, Proceedings, Part I}.\hskip 1em
  plus 0.5em minus 0.4em\relax Springer, 2017, pp. 420--446.

\bibitem{goyal2022tight}
V.~Goyal, Y.~Ishai, and Y.~Song, ``{Tight Bounds on the Randomness Complexity
  of Secure Multiparty Computation},'' in \emph{Advances in Cryptology--CRYPTO
  2022: 42nd Annual International Cryptology Conference, CRYPTO 2022, Santa
  Barbara, CA, USA, August 15--18, 2022, Proceedings, Part IV}.\hskip 1em plus
  0.5em minus 0.4em\relax Springer, 2022, pp. 483--513.

\bibitem{schneider2016lean}
J.~Schneider, ``{Lean and Fast Secure Multi-party Computation: Minimizing
  Communication and Local Computation using a Helper},'' in \emph{Proceedings
  of the 13th International Joint Conference on e-Business and
  Telecommunications}, 2016, pp. 223--230.

\bibitem{Zhao_Sun_Expand}
Y.~Zhao and H.~Sun, ``{Expand-and-Randomize: An Algebraic Approach to Secure
  Computation},'' \emph{Entropy}, vol.~23, no.~11, p. 1461, 2021.

\bibitem{shannon1949}
C.~E. Shannon, ``{Communication Theory of Secrecy Systems},'' \emph{Bell System
  Technical Journal}, vol.~28, no.~4, pp. 656--715, 1949.

\bibitem{selection}
Y.~Zhao and H.~Sun, ``{MDS Variable Generation and Secure Summation with User
  Selection},'' \emph{arXiv preprint arXiv:2211.01220}, 2022.

\bibitem{Zhou_Zhao_Sun}
Z.~Li, Y.~Zhao, and H.~Sun, ``{Weakly Secure Summation with Colluding Users},''
  \emph{arXiv preprint arXiv:2304.09771}, 2023.

\bibitem{li2022arithmetic}
S.~Li and C.~T. Li, ``{Arithmetic Network Coding for Secret Sum Computation},''
  in \emph{2022 IEEE International Symposium on Information Theory
  (ISIT)}.\hskip 1em plus 0.5em minus 0.4em\relax IEEE, 2022, pp. 1034--1039.

\bibitem{vithana2023private}
S.~Vithana, Z.~Wang, and S.~Ulukus, ``{Private Information Retrieval and Its
  Applications: An Introduction, Open Problems, Future Directions},''
  \emph{arXiv preprint arXiv:2304.14397}, 2023.

\bibitem{wang2023fully}
Z.~Wang and S.~Ulukus, ``{Fully Robust Federated Submodel Learning in a
  Distributed Storage System},'' \emph{arXiv preprint arXiv:2306.05402}, 2023.

\end{thebibliography}

\end{document}